\documentclass[12pt,a4paper]{article}

\pdfoutput=1

\usepackage[margin=3.5cm]{geometry}
\usepackage{amsmath,amssymb,graphicx}
\usepackage[dvipsnames]{xcolor}
\usepackage{hyperref}
\usepackage{multicol}
\usepackage{soul}
\usepackage{slashed}
\usepackage{appendix}
\usepackage{changepage}
\usepackage{comment}
\usepackage{authblk}
\usepackage{scalefnt}
\usepackage{setspace}

\newcommand{\red}[1]{\textcolor{red}{#1}}

\numberwithin{equation}{section}
\begin{document}
	

\title{\bf {Schwinger Effect with a \\Transversal Boundary}}

\author{Santiago Christiansen Murguizur, Daniela D'Ascanio,\\Pablo Pisani, Ulises Wainstein Haimovichi}

\affil{Instituto de Física La Plata
	(UNLP/CONICET) Argentina}

\date{}

\maketitle

\thispagestyle{empty}

\begin{abstract}
	{We study the effect of a boundary on pair production by an electric field. In particular, we are interested in the case in which the electric field has a normal component with respect to the boundary. This is essentially more complex than pure tangential fields due to the reflection of created particles against the boundary. In this article we provide the explicit solution for the case of an instantaneous electric pulse which is spatially uniform and normal to a flat surface. We show that pair production exhibits a bulk contribution that accounts for the interference of reflected modes, in addition to a purely boundary term. To conclude, we give an expression for the total particle production due to an irrotational electric pulse in a region delimited by an arbitrary boundary.}
\end{abstract}

\vfill

\noindent\rule{5cm}{.25mm}

\vspace*{-3mm}

\noindent{\footnotesize  santiago.christiansen@fisica.unlp.edu.ar\\[-4mm]
dascanio@fisica.unlp.edu.ar\\[-4mm]
pisani@fisica.unlp.edu.ar\\[-4mm]
ulises2357@hotmail.com}


\pagebreak

\tableofcontents

\vspace{7mm}
\hrule
\vspace{2mm}

\section{Introduction}

{The Schwinger effect is one of the most representative examples of a quantum phenomenon driven by a classical background. Proposed initially \cite{Sauter}
and studied further \cite{H-E,Weisskopf:1936hya}
in the 30's in the context of the Klein paradox, the problem became a cornerstone of QED, being the first non-perturbative effect predicted within its framework. 
The traditional approach for a static, homogeneous electric field involves the evaluation of the imaginary part of the (gauge-invariant) effective action \cite{Schwinger:1951nm}, which gives the asymptotic particle content of the `in' vacuum with respect to the `out' vacuum. A different approach amounts to studying the evolution of the `in' creation and annihilation operators into the `out' ones via a Bogoliubov transformation \cite{Nikishov:1969,Gitman:1977ne}. 
Later, quantum kinetic \cite{Kluger:1991ib,Kluger:1992gb,Smolyansky:1997fc,Schmidt:1998vi} and worldline instantons methods \cite{Affleck:1981bma,Dunne:2005sx} gave new insight into the problem, while also allowing to consider more general background fields. A beautiful chronicle of the technical aspects in the different approaches can be read in \cite{Schubert:2024vft}.

After the foundational analysis of the Sauter step \cite{Sauter2} -- a static, spatially inhomogeneous, localized field which allows for an exact analytical solution -- more realistic settings were studied over the course of time. Beyond the static field case, noticeable examples include alternating electric fields \cite{Brezin:1970xf,Popov:1971}, the temporal Sauter pulse \cite{Narozhnyi:1970uv,Hebenstreit:2010vz,Beltran-Palau:2019bvz}, Gaussian profiles \cite{Popov:1971,Hebenstreit:2009km}, an asymmetric electric field \cite{Breev:2021lpn} and fields varying in both space and time \cite{Schneider:2014mla,Aleksandrov:2016lxd,Aleksandrov:2019ddt}. In the context of worldline instantons, the early works \cite{Dunne:2005sx,Dunne:2006st} address both time- and space-dependent electric fields and show that the instanton trajectories become distorted 
with respect to the circular instanton of a constant field, enhancing (in the time-dependent case) or suppressing (in the space-dependent case) local particle production. {More recently, the Sauter pulse was studied in the context of resurgence theory \cite{Dunne:2022esi}.} An approach for obtaining momentum-resolved quantities was developed in \cite{DegliEsposti:2022yqw}.

Our motivation in the present article is to study the effect of boundaries on the pair production driven by electric fields. In particular, we are interested in configurations in which the electric field is normal to the limiting surface. In this case, reflections from the boundary affect the rate of pair production, hindering the implementation of some of the usual techniques.

The inclusion of boundaries in the analysis of the Schwinger effect was pioneered by \cite{Wang:1988ct}, where the authors consider a static, homogeneous electric field of bounded support to model a capacitor of two plates separated a finite distance. They show how the spatial constraints to the field modify the scattering problem and find that there is a threshold on the distance between plates for particle production to occur. Also, finite-size effects on the particle production probability is shown to be localized at the position of the plates. In \cite{Wong:1994ei} a transversal confinement is introduced, finding an increase on the threshold distance for particle production. Further analysis was carried out in \cite{Gavrilov:2015zem,McGady:2017slr,Breev:2019kug}. In these articles electric fields normal to the boundaries are considered but on these surfaces no restriction is imposed on the quantum fields so produced particles remain free to move across them.

In \cite{Taya:2020pkm} both the background and the created particles are confined within two vibrating plates to study the concomitant interplay between the dynamical Casimir effect and the Schwinger mechanism. However, the electric field is assumed to be parallel to the plates. Pair production on compact manifolds without boundaries was studied in \cite{Brown:2015kgj,Yamada:2024aca}.

Up to our knowledge, the analysis of an electric field normal to a surface that confines the produced particles has only been carried out in \cite{Martin:1988de,Martin:1988gr}. Here, the authors address the case of a static homogeneous field normal to a flat boundary, which could even have constant speed. To simplify the evaluation of the imaginary part of the effective action, the massless case is considered. The result is obtained from the saddle-point approximation up to the single-reflection term of the Balian-Bloch expansion, which yields the contributions proportional to the bulk and boundary volumes.

In the present paper we propose another particular setup which admits an explicit solution. This allows us to study the effects of boundaries transversal to the background electric field in terms of the reflected modes. The key feature -- and more restrictive assumption -- of our model is that the background field is an instantaneous pulse modeled by a Dirac delta function in time. To simplify the setting we place the whole system in the half-line and we remove any spatial dependence from the background field; the spatially uniform pulse is automatically perpendicular to the zero-dimensional boundary. Also for simplicity, we consider scalar QED under Dirichlet boundary conditions on the fluctuating field. Our procedure is straightforward once one selects the appropriate basis modes. Our results allow certain generalizations, such as a finite sequence of electric pulses.

As mentioned above, it is important to recognize the convenient basis of normal modes to expand the quantum field. On spatially uniform time-dependent backgrounds it is useful to exploit translation invariance by expanding the field in canonical-momentum eigenstates. However, in the presence of boundaries these do not satisfy the boundary conditions, so one would use eigenstates of the squared momentum. Nevertheless, when there is a background electric field one needs to use eigenstate of the squared kinetic-momentum.

To construct our solution we rely on the already known result for an instantaneous pulse on the whole real line, which for the sake of clarity we rederive. We also provide a new `geometric' interpretation of the rate of pair production in the line which will offer some insight when analyzing our result on the half-line.

Finally, we also adapt our procedure to the case of a segment of finite length $L$ (two boundaries). The obtained result is less explicit but is apt for numerical evaluation, and we use it to make a comparison, for large $L$, with the result in the half-line. It is not difficult to generalize this case to irrotational instantaneous pulses, i.e.\ with no magnetic counterpart, in regions confined by arbitrary surfaces.

The organization of the article is the following. In Section \ref{timdep} we consider a spatially uniform electric field with an arbitrary time dependence to settle our notation, succinctly review the in-out formalism, and establish Bogoliubov transformation formulae which will be used in the subsequent sections. In Section \ref{sec:line} we consider the particular case of an instantaneous electric field in the line. In Section \ref{sec:half} we introduce a physical boundary by truncating the spatial domain to the half-line. In Section \ref{sec:segment} we consider the finite segment. {In Section \ref{sec:irrot} we present a generalization to the case of an instantaneous and irrotational electric field on an arbitrary compact region.} Finally, general and specific conclusions are drawn in Section \ref{sec:conclu}.


\section{Time Dependent Electric Fields}\label{timdep}

Our specific goal is to study the production of charged scalar particles due to a uniform electric field acting instantaneously in one spatial dimension. We will address, in turn, the following scenarios: the full real line, the half-line and a finite interval. Of course, this setups force the electric field to be perpendicular to the boundary. 

In order to set out our notation and basic formulae, let us first give in this section a short presentation of the in-out procedure in a broader context. In a first instance, we will allow the quantum field to propagate also transversely to a uniform electric field that has, in addition, an arbitrary time dependence. We will shortly specialize to one-dimensional spaces. We thus consider a scalar field $\varphi(x,t)$, with $x=(x_1,x_\perp)$, $x_\perp\in\mathbb{R}^{D-1}$ and $x$ in a one-dimensional manifold $M$ which is either the real line $\mathbb{R}$, the half-line $\mathbb{R^+}$ or the segment $[0,L]$ (this section does not attempt full generality but just avoiding the subsequent repetition of basic formulae). In the last two cases some appropriate boundary conditions are assumed on $\varphi(x,t)$. The scalar field of charge $e$ interacts with an electric field $E(t)$ along the $x_1$-direction that vanishes at asymptotic times, $E(\pm \infty)=0$. We choose the gauge in which the only nonzero component is $A_1=A(t)$, such that $\dot A(t)=E(t)$. The dynamics is given by
\begin{align}\label{dyn}
	\bigg(\partial_t^2-[\partial_1+ieA(t)]^2-\triangle_\perp+m^2\bigg)\varphi(x,t)=0\,.
\end{align}
To write down an appropriate basis of solutions to this equation we first consider the asymptotic spatial operators
\begin{align}
	D^2_\pm=[\partial_1+ieA(\pm\infty)]^2+\triangle_\perp\,,
\end{align}
with the prescribed boundary conditions at $\partial M$. A general eigenstate of $D^2_\pm$ is a combination of plane waves $e^{ikx}$. We denote by $u_k(x)$ the eigenstates of $D_-^2$ and by $v_k(x)$ the eigenstates of $D^2_+$. Both sets of functions are labeled by $k=(k_1,k_\perp)$ with $k_\perp\in\mathbb{R}^{D-1}$ and $k_1$ depending on the choice of $M$ and the boundary conditions. We normalize them as $(u_k,u_{k'})=(v_k,v_{k'})=\delta(k-k')$. The delta is Kronecker or Dirac, accordingly. The eigenvalues are $-k^2=-k_1^2-k_\perp^2$.  We do not need to be more precise at this point.

Next, we denote by $u^\pm_{k}(x,t)$ the solutions to \eqref{dyn} defined by the following behavior at the remote past,
\begin{align}
	u^\pm_{k}(x,t)\underset{t\to-\infty}{\sim}\frac{e^{\mp i\omega_kt}}{\sqrt{2\omega_k}}
	\,u_k(x)\,,
\end{align}
where the oscillation frequencies are
\begin{align}\label{omega}
	\omega_k=+\sqrt{k^2+m^2}\,.
\end{align}
Similarly, we define the solutions $v^\pm_{{k}}(x,t)$ to \eqref{dyn} by their behavior at the remote future,
\begin{align}
	v^\pm_{{k}}(x,t)\underset{t\to\infty}{\sim}\frac{e^{\mp i\omega_kt}}{\sqrt{2\omega_k}}
	\,v_k(x)\,.
\end{align}
 Both set of solutions satisfy the orthogonality conditions
\begin{align}\label{ortho}
	&(u^\pm_k,u^\pm_{k'})_{\mathrm{KG}}=(v^\pm_k,v^\pm_{k'})_{\mathrm{KG}}=\pm\delta(k-k')\,,\nonumber\\ 
	&(u^+_k,u^-_{k'})_{\mathrm{KG}}=(v^+_k,v^-_{k'})_{\mathrm{KG}}=0\,,
\end{align}
with respect to the Klein-Gordon inner product
\begin{align}
	(\phi,\chi)_{\mathrm{KG}}=\int_{M\times \mathbb{R}^{D-1}} d^Dx
	\ \phi^*(x,t)\,i\overset{\leftrightarrow}{\partial_t}\,\chi(x,t)\,,
\end{align}
which -- for $\phi,\chi$ solutions to \eqref{dyn} -- does not depend on the choice of the time slice $t$. At the remote past and future, the wave functions $u^\pm(x,t)$ and $v^\pm(x,t)$, respectively, represent free particles or antiparticles. Therefore, the evolution of $u^\pm(x,t)$ under the influence of the electric field up to $t=\infty$ contains information on the probability of pair production. In fact, at any time, both sets can be related through\footnote{Recall the integral over $k'_1$ should be interpreted as a sum if $M$ is compact.}
\begin{align}\label{trans}
	u^+_{ k}(x,t)&=\int d^Dk'\left[\alpha_{{kk'}}\,v^+_{k'}(x,t)
	+\beta_{kk'}\,v^-_{k'}(x,t)\right]\,,\nonumber\\[2mm]
	u^-_{ k}(x,t)&=\int d^Dk'\left[\beta_{{kk'}}^{*}\,v^+_{k'}(x,t)
	+\alpha_{kk'}^{*}\,v^-_{k'}(x,t)\right]\,,
\end{align}
with appropriate coefficients. The same coefficients appear in the both lines due to the property $(u^+_k,v^\pm_{k'})_{\mathrm{KG}}=-(u_k^-,v^\mp_{k'})_{\mathrm{KG}}^*$, which can be proved by direct computation. In the same sense, the orthogonality relations \eqref{ortho} allows one to compute the Bogoliubov coefficients $\alpha_{kk'},\beta_{kk'}$ in terms of the Klein-Gordon products between $u^\pm_k$ and $v^\pm_{k'}$.

We can now expand the quantized field in either set of solutions,
\begin{align}\label{campo}
	\varphi(x,t)
	&=\int d^Dk
	\left[a_{{k}}\,u^+_{{k}}(x,t)
	+b^\dagger_{{k}}\,u^-_{{k}}(x,t)\right]\nonumber\\[2mm]
	&=\int d^Dk
	\left[\bar a_{{k}}\,v^+_{{k}}(x,t)
	+\bar b^\dagger_{{k}}\,v^-_{{k}}(x,t)\right]
	\,.
\end{align}
Equal-time canonical commutation relations 
lead to Fock algebras $[a_{k},a^\dagger_{k'}]=[b_{k},b^\dagger_{k'}]=[\bar a_{k},\bar a^\dagger_{k'}]=[\bar b_{k},\bar b^\dagger_{k'}]=\delta(k-k')$. For this reason, the coefficients in these expansions are interpreted as creation and annihilation operators of the corresponding `in' or `out' modes.

From the field expansions \eqref{campo}, using \eqref{ortho} and \eqref{trans},
one readily obtains the relation between `in' and `out' creation and annihilation operators,
\begin{align}
	\bar a_{k'}&=\int d^Dk \left(\alpha_{kk'}\,a_{k}+\beta^*_{kk'}\,b^\dagger_{k}\right)\,,\\[2mm]
	\bar b^\dagger_{k'}&=\int d^Dk\left(\beta_{kk'}\,a_{k}+\alpha^*_{kk'}b^\dagger_{k}\right)\,.
\end{align}

As a consequence, the state $|0,{\rm in}\rangle$ (the vacuum at the remote past) represents at the distant future a state whose number of particles in the state created by $\bar a^\dagger_k$ is given by
\begin{align}\label{eq:number}
	\langle 0,{\rm in}|\bar a_k^\dagger \bar a_k|0,{\rm in}\rangle&=
	\int d^Dk'\,d^Dk''\,|\beta_{k'k}|^2\,\langle 0,{\rm in}|{b_{{k'}}b^\dagger_{{k''}}}|0,{\rm in}\rangle
	\nonumber\\[2mm]
	&=
	\int d^Dk'\,|(v^-_{k},u^+_{k'})|^2\,.
\end{align}
In the following sections we will repeatedly use this formula to compute the rate of pair production in the three different settings $M=\mathbb{R}$, $M=\mathbb{R}^+$ and $M=[0,L]$.

\section{Instantaneous Pulse on the Line}\label{sec:line}

We can now apply the results of the previous section to the real line $M=\mathbb{R}$. As shown by \eqref{omega}, the momenta $k_\perp$ transverse to the electric field simply enhances the effective mass of the particles; to maintain notational simplicity, we hereafter drop the extra directions, make $D=1$ and take $x,k\in\mathbb{R}$.

More importantly, we specialize to the case of an instantaneous electric pulse $E(t)=E\,\delta(t)$. Note that the constant $E$ is not strictly the strength of the electric field but rather its impulse $E=\int_{-\infty}^\infty dt\,E\,\delta(t)$. The gauge field is the step function $A(t)=E\,H(t)$.

For $t<0$ there is no background field so the solutions $u^\pm_k(x,t)$ are simply plane waves,
\begin{align}\label{u}
		u^\pm_k(x,t)=\frac1{\sqrt{4\pi\omega_k}}\ e^{\mp i\omega_k t+ikx}
		\qquad{\rm for\ }t<0\,.
\end{align}
Similarly,
\begin{align}\label{v}
		v^\pm_k(x,t)=\frac1{\sqrt{4\pi\omega_k}}\ e^{\mp i\omega_k t+ikx-ieEx}
		\qquad{\rm for\ }t>0\,.
\end{align}
Both set of modes satisfy the normalization \eqref{ortho}. The Klein-Gordon inner product can be conveniently computed at $t=0$, yielding
\begin{align}
		(v^-_{k},u^+_{k'})_{\mathrm{KG}}=\frac{\omega_{k'}-\omega_{k}}{4\pi \sqrt{\omega_k\,\omega_{k'}}}
		\,\delta(k'-k+eE)\,.
\end{align}}
In consequence, the number of produced particles per unit length is straightforwardly given by \eqref{eq:number} as\footnote{The superscript refers to the case without boundaries.}
\begin{align}\label{res:line}
	n^{(0)}_k=\frac{(\omega_{k'}-\omega_{k})^2}{8\pi\,\omega_k\omega_{k'}}\,,
\end{align}
with $k'=k-eE$. This result has already been obtained in \cite{Mahajan:2008ya}. {The analysis for the spinor case can be found in \cite{Ilderton:2021zej}.}

Note that either $u^+_{k'}$ or $u^-_{k'}$ turns after $t=0$ into some combination of $v^+_{k'+eE}$ and $v^-_{k'+eE}$. The wave function is thus a canonical momentum eigenstate, as expected from translation invariance.

Contrary to the static case, this particle creation is a perturbative phenomenon: indeed, in the weak-field limit $eE\ll m$, $n^{(0)}_k\sim E^2$. Note that in this case $n^{(0)}_k$ gives the total particle creation per unit length, so there is no accumulation over time as in the static case. The essential difference between this case and the static one is in the spectral origin of particle creation: while in the static case the constant field carries no real photons and creates particles through quantum tunneling, the delta pulse contains a uniform spectrum of all frequencies and particles are created from the vacuum by absorbing photons from it.

\begin{figure}
	\centering
	\begin{minipage}{.75\linewidth}
		\centering
		\includegraphics[width=\linewidth]{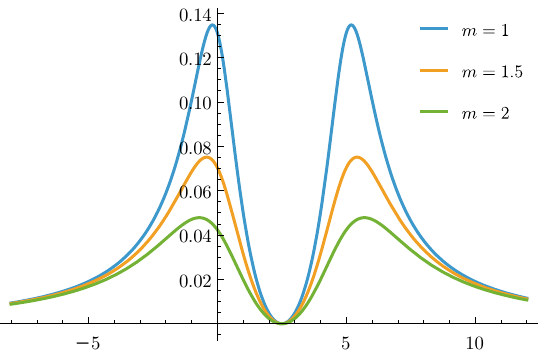}
		\caption{The density of produced pairs as a function of their momentum $k$. We have taken $eE=5$ and $m=1$, $1.5$ and 2. The higher the mass, the smaller the height of the curve.}
		\label{density-recta}
	\end{minipage}
\end{figure}

The result \eqref{res:line} admits an interesting interpretation in the context of relativistic quantum mechanics.
\begin{figure}
	\centering
	\begin{minipage}{.75\linewidth}
		\centering\includegraphics[width=\linewidth]{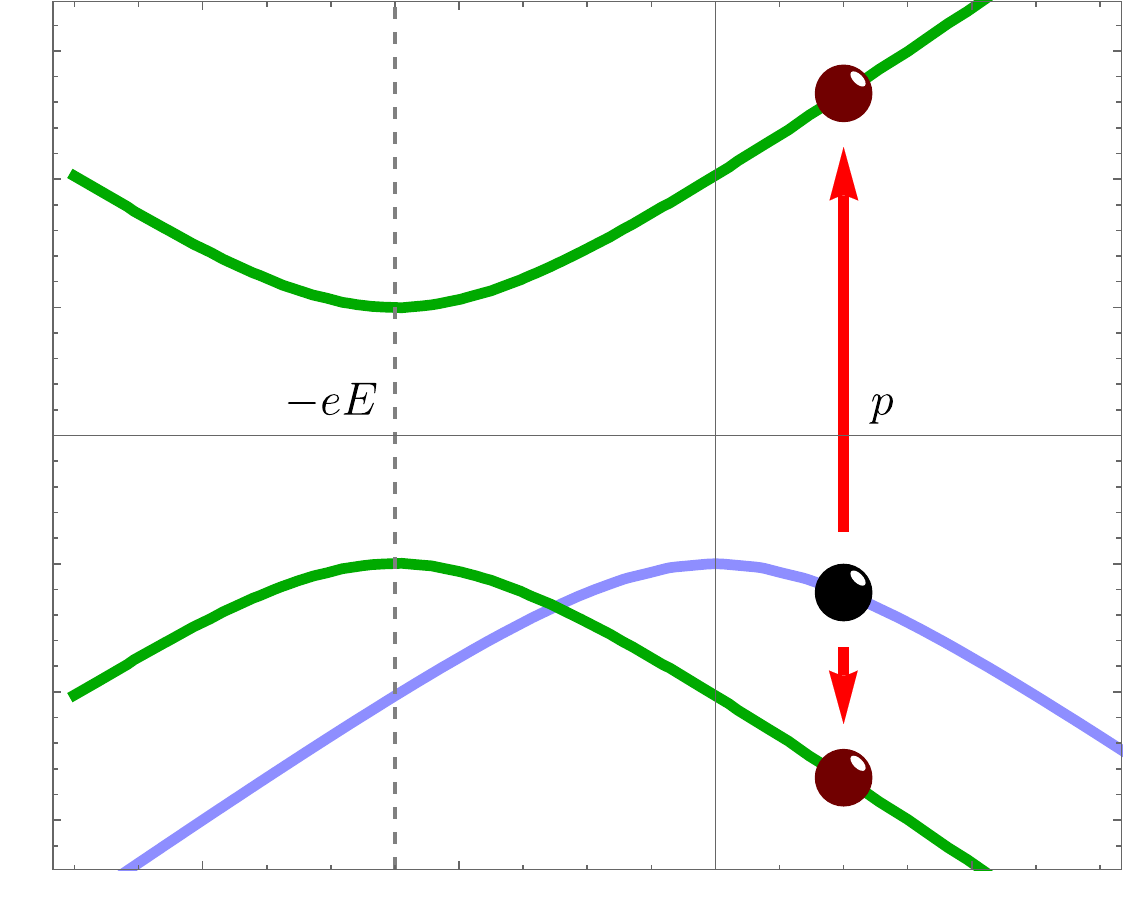}
		\caption{A particle in the Dirac sea can make a transition to another negative energy state or it can surmount the mass gap signaling the creation of a pair particle/antiparticle.}
		\label{hyp}
	\end{minipage}
\end{figure}
Figure \ref{hyp} shows the mass-shell hyperbolae at $t=-\infty$ (blue) and at $t=+\infty$ (green). The horizontal axis represents the canonical momentum; the vertical, the energy. At $t=-\infty$, canonical and kinetic momenta coincide, $p=k'$, and the axis of the mass-shell hyperbola passes through the origin. On the other hand, at $t=+\infty$, they differ, $p=k-eE$, and the hyperbola is shifted to the left an amount $eE$ (we momentarily assume $eE>0$). The black ball represents a particle which is originally in the Dirac sea with canonical momentum $p$. Since the background is invariant under space translations the canonical momentum is preserved. Correspondingly, the kinetic momentum changes an amount $\Delta k=eE>0$, which is the impulse received from the external field. In the figure, conservation of $p$ means that the particle must jump vertically from one hyperbola to the other. There are two possible final states and one would expect the relative transition probabilities to be related with the relative distances to one and the other branch of the final hyperbola. As a matter of fact, one can derive a simple relation between the quotient of these distances and the final density of particles with kinetic momentum $k=p+eE$ given by \eqref{res:line}. With obvious notation,
\begin{align}\label{relhyp}
	\frac{d^2_{(-{\rm blue} \to +{\rm green})}}{d^2_{(-{\rm blue} \to -{\rm green})}}
	=\left(\frac{\sqrt{(p+eE)^2+m^2}+\sqrt{p^2+m^2}}{-\sqrt{(p+eE)^2+m^2}+\sqrt{p^2+m^2}}\right)^2
	=1+\frac{1}{2\pi n^{(0)}_k}\,.
\end{align}
Figure \ref{density-recta} displays the behavior of $n^{(0)}_k$ as a function of $k$.
Relation \eqref{relhyp} allows us to give an explanation of the main features of this plot, namely, two maxima at $eE/2\pm\sqrt{(eE/2)^2+m^2}$ (close to $k=0$ and $k=eE$ for small masses) and a zero at the midpoint $k=eE/2$. In fact, as is clear from Figure \ref{hyp}, the distance $d_{(-{\rm blue} \to +{\rm green})}$ from the negative energy initial branch to the positive energy final branch is almost constant in the interval delimited by the vertices of the hyperbolae. Beyond the vertices this distance increases considerably (specially for small mass) because either one or the other hyperbola changes direction abruptly. For this reason, $n^{(0)}_k$ presents a sharp decrease for $k<0$ and $k>eE$. However, between the vertices $n^{(0)}_k$ is not constant but decreases to zero at the mid-point, due to the denominator $d_{(-{\rm blue} \to -{\rm green})}$, which vanishes there.

The absence of created particles at $k=eE/2$ can also be explained on more physical grounds, in terms of the kinetic momentum gained by the field modes. The electric pulse instantaneously transfers a kinetic momentum $\Delta k = eE$ to the scalar particles. This momentum kick changes $k=-eE/2$ into $k=eE/2$, so it only reverses the direction of motion of a particle in the Dirac sea and does not change its energy. In other words, for $k'=-eE/2$, $k=k'+eE=eE/2$, $u^-_{k'}$ and $v^+_k$ coincide and there is no probability of pair creation. In general, the Hamiltonian undergoes a sudden change and the `in' vacuum results in a superposition of different states of the new Hamiltonian, which amounts to a nontrivial particle density after the kick.

The expression for $n^{(0)}_k$ can be rewritten in terms of dimensionless momentum $\kappa=k/m$ and electric field strength $g=eE/m$ as
\begin{align}\label{keldysh}
	n^{(0)}_\kappa
	=\frac{1}{8\pi}\frac{\left(\sqrt{(\kappa+g)^2+1}-\sqrt{\kappa^2+1}\right)^2}{\sqrt{\kappa^2+1}\sqrt{(\kappa+g)^2+1}},
\end{align}
highlighting the relevance of the Keldysh parameter $\gamma=1/g$. {The pulse $E\,\delta(t)$ can be viewed as the short-width and large-height limit $\tau\to 0,E_0\to\infty$ (with $2E_0\tau=E$) of the Sauter pulse $E(t)=E_0\,{\rm sech}^2(t/\tau)$, whose rate of pair production reduces to \eqref{keldysh}.}

{For large values of $k$, the density $n_k^{(0)}$ decays as $1/k^2$. This means that the total density of particles created is finite. Indeed, the integral of $n_k^{(0)}$ can be explicitly solved in terms of elliptic integrals, giving}
\begin{align}
	n^{(0)}=\frac{m}{\pi}
	\sqrt{1+\frac{g^2}{4}}\left[ K\left(\sqrt{\frac{g^2/4}{1+g^2/4}}\right)-E\left(\sqrt{\frac{g^2/4}{1+g^2/4}}\right)\right]\,.
\end{align}
{Since for large $k$ the energy scales as $k$, there is an infinite energy transfer from the background to the quantum field.}

\section{The Electric Pulse on the Half-Line}\label{sec:half}

In this section we introduce a boundary by considering the instantaneous field in the half-line $M=\mathbb{R}^+$.
We impose Dirichlet boundary conditions on the scalar field at $x=0$. This affects the spatial dependence of the solutions $u_k^\pm(x,t)$ and $v_k^\pm(x,t)$, which arise from the appropriate combination of $\eqref{u}$ and $\eqref{v}$,
\begin{align}
	u^\pm_k(x,t)&=\frac{e^{\mp i\omega_k t}}{\sqrt{\pi\omega_k}}\,\sin{kx}
	& \mathrm{for}\ t<0\,,\\[2mm]
	v^\pm_k(x,t)&=\frac{e^{\mp i\omega_k t}}{\sqrt{\pi\omega_k}}\,e^{-ieEx}\sin{kx}
	& \mathrm{for}\ t>0\,,
\end{align}
with $k>0$.
As before, any solution can be written as a superposition of the `in' or the `out' modes, and both sets are related by \eqref{trans} through the Bogoliubov coefficients $\alpha_{kk'}$ and $\beta_{kk'}$. We will use equation \eqref{eq:number} to obtain the density of particle created at a given momentum mode. We first compute the coefficient $\beta_{kk'}$:
\begin{align}\label{beta1}
	\beta_{kk'}= -(v_{k'}^-,u_k^+)=
	\frac{\omega_{k'}-\omega_k}{\pi\sqrt{\omega_k\omega_{k'}}}
	\int_0^\infty dx
	\ \sin{(kx)}\sin{(k'x)}\,e^{ieEx}\,,
\end{align}
where
\begin{align}\label{beta1p}
	\int_0^\infty dx
	\ \sin{(kx)}\sin{(k'x)}\,e^{ieEx}
	&=\frac i4
	\left(\frac{1}{k-k'+eE+i\epsilon}-\frac{1}{k-k'-eE-i\epsilon}+\mbox{}\right.\nonumber\\[2mm]
	&\left.\mbox{}-\frac{1}{k+k'+eE+i\epsilon}+\frac{1}{k+k'-eE-i\epsilon}\right)\,.
\end{align}
The regularization can be interpreted as a cutoff factor $e^{-\epsilon x}$ in the functions $u^\pm_k,v^\pm_k$. The regulator $\epsilon>0$ is then related to the volume $L$ of the half-line through, e.g.,
\begin{align}
	(u^+_k,u^+_k)_{\mathrm{KG}}&=\frac{2}{\pi} \int_0^\infty dx
	\ \sin^2{kx}\,e^{-2\epsilon x}=\frac{1}{2\pi\epsilon}\nonumber\\[2mm]
	&=\delta(0)=\frac{L}{\pi}\,.
\end{align}
The second line arises from \eqref{ortho}. This shows that the infrared regulator can be interpreted as $\epsilon=1/2L$.

We must then integrate
\begin{align}\label{alpha}
	|\beta_{kk'}|^2\!=\!
	\frac{(\omega_{k'}-\omega_k)^2}{16\pi^2\omega_k\omega_{k'}}
	\!\left|\frac{1}{k'\!-\!k_+\!\!-\!i\epsilon}\!+\!\frac{1}{k'\!+\!k_+\!\!+\!i\epsilon}\!-\!\frac{1}{k'\!-\!k_-\!\!+\!i\epsilon}\!-\!\frac{1}{k'\!+\!k_-\!\!-\!i\epsilon}\right|^2\!,
\end{align}
where we have introduced the kinematic parameters $k_\pm=k\pm eE$.
Using the partial fractions decomposition
\begin{align}\label{uli}
	\frac{1}{(k'-A)(k'-B)}=\frac{1}{A-B}\left(\frac{1}{k'-A}-\frac{1}{k'-B}\right)
\end{align}
we can break down $|\beta_{kk'}|^2$ in \eqref{alpha} into separate terms, each of them yielding an integral of the type
\begin{align}\label{defint}
	I(A)=\int_0^\Lambda dk'
	\ \frac{(\omega_{k'}-\omega_k)^2}{\omega_k\omega_{k'}}
	\ \frac{1}{k'-A}\,.
\end{align}
Here $A=a+i\epsilon$, with $a=\pm k_\pm$. Eventually, we will also need $I(A^*)=[I(A)]^*$. The regulator $\Lambda$ is introduced to ensure the finiteness of the integral for each of term separately. Power counting shows that $I(A)$ contains $\Lambda$- and $\log{\Lambda}$-terms which, ultimately, cancel so that, as we collect all contributions, we will safely take $\Lambda\to\infty$. Of course, we will neglect any term $O(\Lambda^{-1})$.

As regards the limit $\epsilon\to 0^+$, on physical grounds we expect a contribution to the number of created pairs proportional to the volume $L=1/2\epsilon$. One can readily see that this divergence comes from the four squared terms in \eqref{alpha}, which lead to $B=A^*$ in \eqref{uli} and, consequently, $A-B\sim \epsilon$. Therefore, we will retain $O(\epsilon^{0})$ as well as $O(\epsilon^1)$ in the computation of $I(A)$. In this way we will distinguish bulk from boundary contributions.

The definite integral $I(A)$ can be computed from the primitive
\begin{align}\label{pri}
	\int dk'
	\ \frac{(\omega_{k'}-\omega_k)^2}{\omega_k\omega_{k'}}
	\ \frac{1}{k'-A}
	&=\frac{\omega_{k'}}{\omega_k}
	+\frac{A}{2\omega_k}\,\log{\left(\frac{\omega_{k'}+k'}{\omega_{k'}-k'}\right)}
	-2\log{(k'-A)}
	+\mbox{}\nonumber\\[2mm]
	&\mbox{}-\frac12\left(\frac{\omega_{A}}{\omega_k}+\frac{\omega_{k}}{\omega_A}\right)
	\log{\left(\frac{{\omega_{A}\omega_{k'}}+Ak'+m^2}{{\omega_{A}\omega_{k'}}-Ak'-m^2}\right)}
	\,,
\end{align}
but some care must be taken in the choice of the Riemann sheet in order to preserve the analyticity of the three logarithmic terms as $k'$ runs in the interval $(0,\Lambda)$. The argument of the first logarithm is always positive so its phase can be simply chosen as zero. The second logarithm does not involve any subtleties either, and for later convenience we choose the cut in the negative real axis. The third logarithm, in the second line, requires a more meticulous analysis. For small $\epsilon$ and large $\Lambda$ its numerator has the following endpoints in the complex plane:
\begin{align}\label{endpt}
	\left.\omega_{A}\omega_{k'}+Ak'+m^2\right|_{k'=0}
	&\simeq m\,(\omega_a+m)\left(1+i\,\frac{a}{\omega_a(\omega_a+m)}\,\epsilon\right)\,,
	\nonumber\\[2mm]
	\left.\omega_{A}\omega_{k'}+Ak'+m^2\right|_{k'=\Lambda}
	&\simeq\left(\omega_a+a\right)\Lambda
	\left(1+i\,\frac{\epsilon}{\omega_a}\right)\,.
\end{align}
For $a>0$ both endpoints lie in the upper half-plane; in fact, this numerator never crosses the real axis, for $k'\in(0,\Lambda)$. In consequence, we can safely define the complex phases within the interval $(0,2\pi)$. For $a<0$ the endpoints \eqref{endpt} lie instead at different sides of the real axis, and the numerator crosses the real axis (for $k'=-a$) at a positive abscissa $2m^2 +O(\epsilon^2)$. As a consequence, for negative $a$ one must choose the logarithmic cut such that the phases lie at the interval $(-\pi,\pi)$.

Let us now analyze the denominator of the logarithm in the second line of \eqref{pri} for $k'\in(0,\Lambda)$:
\begin{align}
	\left.\omega_{A}\omega_{k'}-Ak'-m^2\ \right|_{k'=0}
	&\simeq m\,(\omega_A-m)\left(1+i\,\frac{A}{\omega_A(\omega_A-m)}\,\epsilon\right)\,,
	\nonumber\\[2mm]
	\left.\omega_{A}\omega_{k'}-Ak'-m^2\ \right|_{k'=\Lambda}
	&\simeq\left(\omega_A-A\right)\Lambda
	\left(1-i\,\frac{\epsilon}{\omega_A}\right)\,.
\end{align}
For $a>0$ the endpoints lie at different sides of the real axis and the denominator crosses the negative real axis (for $k'=a$) at $-m^2\epsilon^2/(2\omega_\alpha^2)+O(\epsilon^4)$. For this reason we choose the cut along $[0,\infty)$, as we did for the numerator. For $a<0$ the denominator does not intersect the real axis so we can maintain the cut along $(-\infty,0]$.

Having appropriately defined the complex phases -- in $(0,2\pi)$ for positive $a$ and in $(-\pi,\pi)$ for negative $a$ -- we can now evaluate the definite integral \eqref{defint}. After some algebra and many cancellations we obtain
\begin{align}\label{ia}
	I(A)&=
	\frac{\Lambda-m}{\omega_k}
	+\frac{a+i\epsilon´}{\omega_k}\,\log{\left(\frac{2\Lambda}{m}\right)}
	-2\,\log{\left(\frac{\Lambda}{|a|}\right)}
	+\mbox{}\nonumber\\[2mm]
	&\mbox{}+i\pi\left(\frac{\omega_{a}}{\omega_k}+\frac{\omega_{k}}{\omega_a}-2\right)
	H(a)
	-\pi\epsilon\frac{a}{\omega_{a}^2}
	\left(\frac{\omega_{a}}{\omega_{k}}-\frac{\omega_{k}}{\omega_{a}}\right)
	H(a)
	+\mbox{}\nonumber\\[2mm]
	&\mbox{}+\frac12\left(\frac{\omega_{a}}{\omega_k}+\frac{\omega_{k}}{\omega_a}\right)
	\log{\left[\left(\frac{\omega_{a}+m}{\omega_{a}-m}\right)
		\left(\frac{\omega_{a}-a}{\omega_{a}+a}\right)\right]}
	+\mbox{}\nonumber\\[2mm]
	&\mbox{}+\frac12\,i\epsilon\ \frac{a}{\omega_{a}^2}
	\left(\frac{\omega_{a}}{\omega_{k}}-\frac{\omega_{k}}{\omega_{a}}\right)
	\log{\left[\left(\frac{\omega_{a}+m}{\omega_{a}-m}\right)
		\left(\frac{\omega_{a}-a}{\omega_{a}+a}\right)\right]}+\mbox{}\nonumber\\[2mm]
	&\mbox{}-i\epsilon\ {\frac{1}{\omega_{a}}}
	\left(\frac{\omega_{a}}{\omega_k}+\frac{\omega_{k}}{\omega_a}\right)
	\left(1+\frac{m}{a}\right) +\frac{2i\epsilon}{a} \,.
\end{align}

Repeated application of this formula to each term arising from \eqref{alpha} and \eqref{uli} allows us to write the particle number density in the half-line \eqref{eq:number} as\footnote{The superscript stands for the number of boundaries of $\mathbb{R}^+$.}
\begin{align}
		n_k^{(1)}=\frac{1}{16\pi^2}&\text{Re}\bigg[I(k_++i\epsilon)\left(\frac{1}{i\epsilon}+\frac{1}{k_+}-\frac{1}{eE}-\frac{1}{k}\right)+\mbox\nonumber\\[2mm]
		&+I(-k_++i\epsilon)\left(\frac{1}{i\epsilon}-\frac{1}{k_+}+\frac{1}{eE}+\frac{1}{k}\right)+\mbox\nonumber\\[2mm]
		&+I(k_-+i\epsilon)\left(\frac{1}{i\epsilon}+\frac{1}{k_-}+\frac{1}{eE}-\frac{1}{k}\right)+\mbox\nonumber\\[2mm]
		&+I(-k_-+i\epsilon)\left(\frac{1}{i\epsilon}-\frac{1}{k_-}-\frac{1}{eE}+\frac{1}{k}\right)\bigg]\,.
\end{align}
One can readily see that in this expression all divergent terms in $\Lambda$ (linear and logarithmic) cancel. Using \eqref{ia} we finally get
\begin{align}\label{res:half}
	&{n^{(1)}_k} =\frac{1}{8\pi}
	{\left[
	\frac{(\omega_+-\omega_k)^2}{\omega_+\omega_k}+\frac{(\omega_k-\omega_-)^2}{\omega_k\omega_-}
	\right]}
	+\mbox{}\nonumber\\[2mm]
	&\mbox{}+\frac{1}{16\pi^2L}\left[\frac{k_+}{\omega_{+}^2}
	\left(\frac{\omega_{+}}{\omega_{k}}-\frac{\omega_{k}}{\omega_{+}}\right)
	\log{\left(\frac{\omega_{+}-k_+}{\omega_{+}+k_+}\right)}
	+\frac{k_-}{\omega_{-}^2}
	\left(\frac{\omega_{-}}{\omega_{k}}-\frac{\omega_{k}}{\omega_{-}}\right)
	\log{\left(\frac{\omega_{-}-k_-}{\omega_{-}+k_-}\right)}\right.+\mbox{}\nonumber\\[2mm]
	&\phantom{+\frac{1}{16\pi^2\red{L^2}}\left[\right]}-{\frac{2}{\omega_{+}}}
	\left(\frac{\omega_{+}}{\omega_k}+\frac{\omega_{k}}{\omega_+}\right)
	-{\frac{2}{\omega_{-}}}
	\left(\frac{\omega_{-}}{\omega_k}+\frac{\omega_{k}}{\omega_-}\right)
	+\mbox{}\nonumber\\[2mm]
	&\phantom{+\frac{1}{16\pi^2\red{L^2}}\left[\right]}-\frac{3k_+^2+k_-^2}{k_+(k_+^2-k_-^2)}
	\left(\frac{\omega_{+}}{\omega_k}+\frac{\omega_{k}}{\omega_+}\right)
	\log{\left(\frac{\omega_{+}-k_+}{\omega_{+}+k_+}\right)}
	+\mbox{}\nonumber\\[2mm]
	&\left.
	\phantom{+\frac{1}{16\pi^2\red{L^2}}\left[\right]}
	{-\frac{3k_-^2+k_+^2}{k_-(k_-^2-k_+^2)}}
	\left(\frac{\omega_{-}}{\omega_k}+\frac{\omega_{k}}{\omega_-}\right)
	\log{\left(\frac{\omega_{-}-k_-}{\omega_{-}+k_-}\right)}\right]
	\,.
\end{align}
Unlike the case in the whole real line, $n^{(1)}_k$ is -- as expected -- symmetric under the interchange $eE\leftrightarrow -eE$. Hereafter, we take $eE>0$.

\begin{figure}
	\centering
	\begin{minipage}{.75\linewidth}
		\centering
		\includegraphics[width=\linewidth]{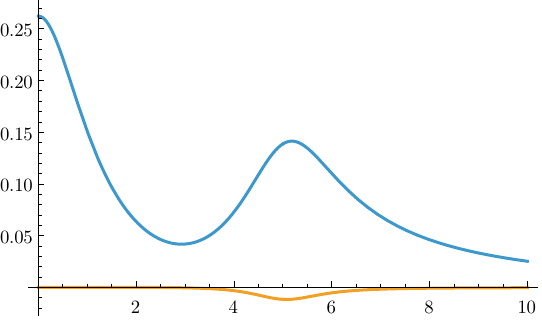}
		\caption{The density of produced pairs in the half-line as a function of their mode number $k$. The orange curve displays the purely boundary contribution. We have taken $eE=5$ and $L=10$ in units where $m=1$.}
		\label{delta_half}
	\end{minipage}\hspace{10mm}
\end{figure}

The behavior of the particle density with the mode number $k$ is portrayed in Figure \ref{delta_half}.
The first line in \eqref{res:half} is the volume term, corresponding to the bulk particle production. As one would expect, it amounts to a symmetrization around the origin of the result \eqref{res:line} on the whole line, $n^{{(1)},\mathrm{bulk}}_k=n^{(0)}_k+n^{(0)}_{-k}$. This is a momentum-space manifestation of the method of images: in the line, the electric field accelerates particles in one direction and antiparticles in the other, but in the half-line the boundary acts as a perfectly reflection wall, and the modes are now standing waves. The bulk density for the standing wave mode $k$ thus accounts for particles created moving to the right and particles created moving to the left that bounce off the wall.

\begin{figure}
	\centering
	\begin{minipage}{.75\linewidth}
		\centering
		\includegraphics[width=\linewidth]{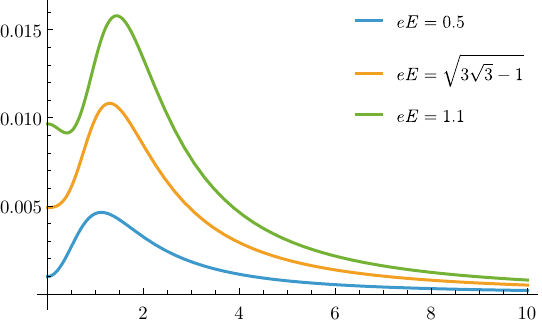}
		\caption{The bulk density of produced pairs in the half-line as a function of their mode number $k$. We have taken $m=1$ and $eE=0.5$ (blue curve), $E=\sqrt{\sqrt{3}-1}$ (orange curve) and $E=1.1$ (green curve).}
		\label{delta_half-regimes}
	\end{minipage}\hspace{10mm}
\end{figure}

As the symmetrization of $n^{(0)}_k$, the bulk density $n^{(1),\mathrm{bulk}}_k$ has a critical point at $k=0$. Recall that for small fields $n^{(0)}_k$ has its maxima at $k\simeq \pm m$ and the minimum at $k\simeq 0$. Symmetrization results in a minimum of $n^{(1),\mathrm{bulk}}_k$ at $k=0$ and only one maximum at $k\simeq m$. On the other hand, for large fields $n^{(0)}_k$ has one maximum near $k=0$ and the second one near $k=eE$. Symmetrization implies that $n^{(1),\mathrm{bulk}}_k$ has a maximum at $k=0$ and another maximum close to $k=eE$. These two regimes, depicted in Figure \ref{delta_half-regimes}, are separated by the critical value $eE_c=\sqrt{\sqrt{3}-1}\,m\simeq 0.8556\,m$.

In the case of very strong fields $eE\gg m$, the right maximum of $n^{(0)}_{k}$ approaches $k=eE$ and the left maximum is close to $k=0$, so it has a peak of particles that absorb almost all the momentum strength of the kick and a peak of particles created almost at rest. This allows to think of the two field regimes in another way: fields smaller than the critical field are not strong enough to completely separate the two populations of particles, resulting in a broad spread where the most likely momentum is not zero, and fields higher than the critical field push the particles that absorb the most of the momentum kick to the right, creating an outer maximum for $n^{(1),\mathrm{bulk}}$. In the critical case, the inner peaks of $n^{(0)}_k$ and $n^{(0)}_{-k}$ merge at the origin.

The remaining terms $n^{(1),\mathrm{bdy}}_k$ correspond to corrections to the bulk result induced by the existence of the boundary. This contribution is independent of the size of the space. The symmetry under the interchange $k_+\longleftrightarrow k_-$ echoes the symmetry of the half-line. The apparent infrared divergence of the last two terms is cured by this symmetry; $n^{(1),\mathrm{bdy}}_k$ remains finite (actually vanishes) at $k=0$, leaving the bulk density as the only contribution.

Numerical exploration shows that $n^{(1),\mathrm{bdy}}_k$ is mostly negative and has a local minimum near the outer maximum of $n^{(1),\mathrm{bulk}}_k$. In fact, $n^{(1),\mathrm{bdy}}_k$ is zero at $k=0$ and $n^{(1),\mathrm{bdy}}_k\to 0^-$ as $k\to\infty$ so it must have a minimum where it takes a negative value. We conclude that, as expected, the effect of Dirichlet boundary conditions is to obstruct particle creation.

One can see the rapidity parameter to appear in the logarithms: 
\begin{align}
	\log\left(\frac{\omega_k-k}{\omega_k+k}\right)
	=-2\operatorname{arctanh}\left(\frac{k}{\omega_k}\right)\equiv-2\eta_k\,,
\end{align}
where $v_k = k/\omega_k$ is the relativistic velocity of the mode $k$, which corresponds to the boost relating the boundary and the particle rest frames.

The large $k$ limit of \eqref{res:half} shows $n^{(1),{\rm bulk}}\sim 1/k^2$ and $n^{(1),{\rm bdy}}\sim 1/k^3$. As a consequence, the number of created particles is finite but the energy absorbed from the electric field by the bulk contribution diverges.

\section{The Segment}\label{sec:segment}

In this section we apply the electric pulse in the interval $M=[0,L]$. For comparison, we impose again Dirichlet boundary conditions at the endpoints. The appropriately normalized `in' and `out' solutions now read
\begin{align}
	u_n^\pm(x,t)&=\frac{1}{\sqrt{\omega_{k_n} L}}\,\sin{\left(\frac{\pi n}{L}\,x\right)}
	\,e^{\mp i\omega_{k_n} t}& {\rm for\ }t<0\,,\nonumber\\[2mm]
	v_n^\pm(x,t)&=\frac{1}{\sqrt{\omega_{k_n} L}}\,\sin{\left(\frac{\pi n}{L}\,x\right)}
	\,e^{ieEx}
	\,e^{\mp i\omega_{k_n} t}& {\rm for\ }t>0\,.
\end{align}
The kinetic momenta become discrete, $k_n=\pi n/L$, with $n=1,2,\ldots$\,. To apply the Bogoliubov formalism described above we replace Dirac deltas with Kronecker deltas and integrals with sums. The density of particles created in the $n$-th momentum mode is thus
\begin{align}\label{eq:density-interval}
	{n_n^{(2)}=\frac 1{L}\,}\langle 0,{\rm in}|\bar a_n^\dagger\bar a_n|0,{\rm in}\rangle
	={\frac 1{L}\,}\sum_{p=1}^\infty
	\ |(v^+_n,u^-_p)|^2\,,
\end{align}
The scalar product in the sum gives
\begin{align}
	(v^+_n,u^-_p)
	=\frac i4\left(\sqrt{\frac{\omega_{k_n}}{\omega_{k_p}}}-\sqrt{\frac{\omega_{k_p}}{\omega_{k_n}}}\right)
	\sum_{r,s=\pm}
	\ rs\,\frac{(-1)^{n+p}e^{-ieEL}-1}{\pi(rn+sp)-eEL}
	\,,
\end{align}
which is the discretized version of \eqref{beta1} and \eqref{beta1p}. Its squared modulus yields
\begin{align}\label{eq:interv-amplitude}
	&|(v^+_n,u^-_p)|^2
	=
	\frac{(4eE)^2}{L^2}
	{\frac{(\omega_{k_n}-\omega_{k_p})^2}{\omega_{k_p}\omega_{k_n}}}\ k_n^2\,k_p^2\times\mbox{} \nonumber\\[2mm]
	& \mbox{}\times \frac{\sin^2{[\tfrac{L}2(k_n+k_p-eE)]}}
	{[(k_n+k_p+eE)(k_n+k_p-eE)(k_n-k_p+eE)(k_n-k_p-eE)]^2}
	\,.
\end{align}
Contrary to the continuous case, due to discretization, the apparent singularities at $k_n = \pm e E  \pm k_p$ are directly canceled by the sine function in the numerator. {We note the presence of a frequency mismatch factor $(\omega_{k_n}-\omega_{k_p})^2/\omega_{k_n}\omega_{k_p}$, which makes the diagonal contribution to particle production vanish. This will be a feature of the more general result in Section \ref{sec:irrot}.}

\begin{figure}
	\centering\begin{minipage}{.75\linewidth}
		\centering
		\includegraphics[width=\linewidth]{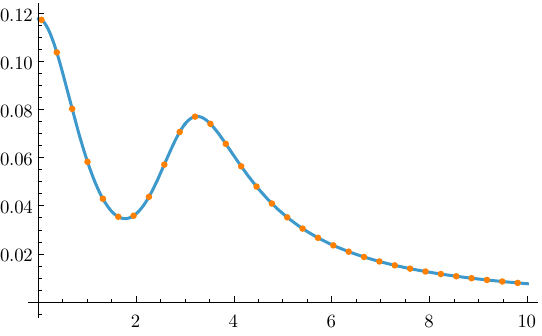}
		\caption{{Density of} produced pairs as a function of the {mode momentum number}. Here $L=50$ and $eE=3$ in units where $m=1$. {For better visualization, we display only the momenta $k_n$ where $n$ is a multiple of 5.} The blue curve corresponds to the infinite half-line (with a doubled boundary term).}
		\label{comp1}
		\end{minipage}
\end{figure}

For large $p$, the $p$-th term in this expression decays as $p^{-5}$, so the series is absolutely convergent. The sum in \eqref{eq:density-interval} gives the density of particle production for each $n$-th mode, which is represented by the dots in Figure \ref{comp1}.
	
To leading order, the continuum limit $L\to\infty$ readily gives the bulk term of the half-line density:
taking into account the momentum spacing, the sum $\sum_{p=1}^\infty$ becomes the integral $L/\pi\int_0^\infty dp$, and as the integrated momentum reaches the zeros of the denominator the limit introduces a delta function as $\lim_{L\rightarrow\infty}\sin^2(Lk/2)/k^2=\pi L\delta(k)/2$, giving
\begin{align}
	\lim_{L\rightarrow\infty}\ \frac L\pi\,n_{n}^{(2)}= \frac1{8\pi} \left( \frac{\omega_{k+eE}}{\omega_k} + \frac{\omega_k}{\omega_{k+eE}} +\frac{\omega_{|k-eE|}}{\omega_k}+\frac{\omega_k}{\omega_{|k-eE|}} -4 \right)= n_k^{(1),\mathrm{bulk}}\,.
\end{align}
The computation of the subleading term $O(L^{-1})$ is more involved but, after a numerical exploration, one finds, as expected, twice the boundary term of the half-line density. The full agreement is shown in Figure \ref{comp1}.

Numerical analysis also indicates that $n^{(2)}_n\sim 1/n^2$ for large $n$ (as in our previous examples). Therefore, the total density of created particles is finite and given by the double sum
\begin{align}
n^{(2)} = \sum_{n=1}^\infty n_n^{(2)} = \frac1{L}\sum_{n,p=1}^\infty
|(v^+_n,u^-_p)|^2\,.
\end{align}
Once more, the total energy density generated by the field is infinite.

\section{Irrotational Pulse on an Arbitrary Region}\label{sec:irrot}

Before concluding, we write down a generalization of the computations of the previous section to an arbitrary compact region $M\subset\mathbb{R}^D$ with boundary $\partial M$ and a non homogeneous electric pulse. On the line, translation invariance allows the problem to be formulated entirely in terms of momentum modes. As we have seen, when boundaries are present, momentum is no longer an appropriate quantum number and particle production is instead characterized by the mixing of the normal modes of the system. We formulate the instantaneous field problem in terms of the spectral decomposition of the Laplacian with some boundary conditions at $\partial M$.

We consider a charged scalar field $\varphi(t,x)$, with $x\in M$, subject to an instantaneous and irrotational electric field $E_i(t,x)=-\partial_i\phi(x)\,\delta(t)$, for some smooth $\phi(x)$. This electric pulse is compatible with the absence of magnetic field. We choose the gauge $A_i=-\partial_i\phi(x) \,H(t)$.

Solutions $u^\pm_\lambda(x,t),v^\pm_\lambda(x,t)$ to the Klein-Gordon equation with the expected behavior at the remote past and future are defined by
\begin{align}
	u_\lambda^\pm(t,x)
	&=\frac{1}{\sqrt{2\omega_\lambda}}\, e^{\mp i\omega_\lambda t}\,e_\lambda(x)
	\qquad {\rm for\ }t<0\,,
	\\[2mm]
	v_\lambda^\pm(t,x)
	&=\frac{1}{\sqrt{2\omega_\lambda}}\, e^{\mp i\omega_\lambda t}\,e_\lambda(x)\,e^{ie\phi(x)}
	\qquad {\rm for\ }t>0\,,
\end{align}
where $e_\lambda(x)$ are Laplacian eigenstates,
\begin{align}
	-\triangle\,e_\lambda(x)=\lambda\,e_\lambda(x)\,,
\end{align}
with eigenvalue $\lambda$, that satisfy the assumed boundary conditions at $\partial M$. The frequency of the normal modes is $\omega_\lambda=\sqrt{\lambda+m^2}$. If the normalization is chosen such that $(e_\lambda,e_{\lambda'})=\delta_{\lambda\lambda'}$ then $u^\pm_\lambda(x,t),v^\pm_\lambda(x,t)$ satisfy normalization \eqref{ortho}.

At any time $u_\lambda^+(t,x)$ is a combination of all $v_\lambda^\pm(t,x)$,
\begin{align}
	u_\lambda^+(t,x)=\sum_{\lambda'}
	\left(\alpha_{\lambda\lambda'}\,v_{\lambda'}^+(t,x)
	+\beta_{\lambda\lambda'}\,v_{\lambda'}^-(t,x)\right)\,.
\end{align}
with,
\begin{align}
	\beta_{\lambda\lambda'}=
	\frac12\left(\frac{\sqrt\omega_{\lambda'}}{\sqrt\omega_{\lambda}}
	-\frac{\sqrt\omega_\lambda}{\sqrt\omega_{\lambda'}}\right)
	(e_{\lambda'}(x),e_\lambda(x)\,e^{-ie\phi(x)}).
\end{align}
Here we identify the frequency mismatch factor, which represents the effect of the electric field on the scalar vacuum. The effect of boundaries is encoded in the matrix elements of the gauge phase $e^{-ie\phi(x)}$.

By adding up the contributions of all modes $\lambda$ and $\lambda'$ one obtains the number of particles {in the mode $\lambda$} created by the instantaneous pulse,
\begin{align}\label{eq:number-irrot-discrete}
	N_{\lambda}=\sum_{\lambda'}|\beta_{\lambda\lambda'}|^2
	=\sum_{\lambda'}
	\frac{(\omega_\lambda-\omega_{\lambda'})^2}{4\,{\omega_{\lambda'}}{\omega_{\lambda}}}
	\left|(e_{\lambda'}(x),e_\lambda(x)\,e^{-ie\phi(x)})\right|^2
	\,.
\end{align}

\begin{figure}
	\centering
			\begin{minipage}{.75\linewidth}
		\centering
		\includegraphics[width=\linewidth]{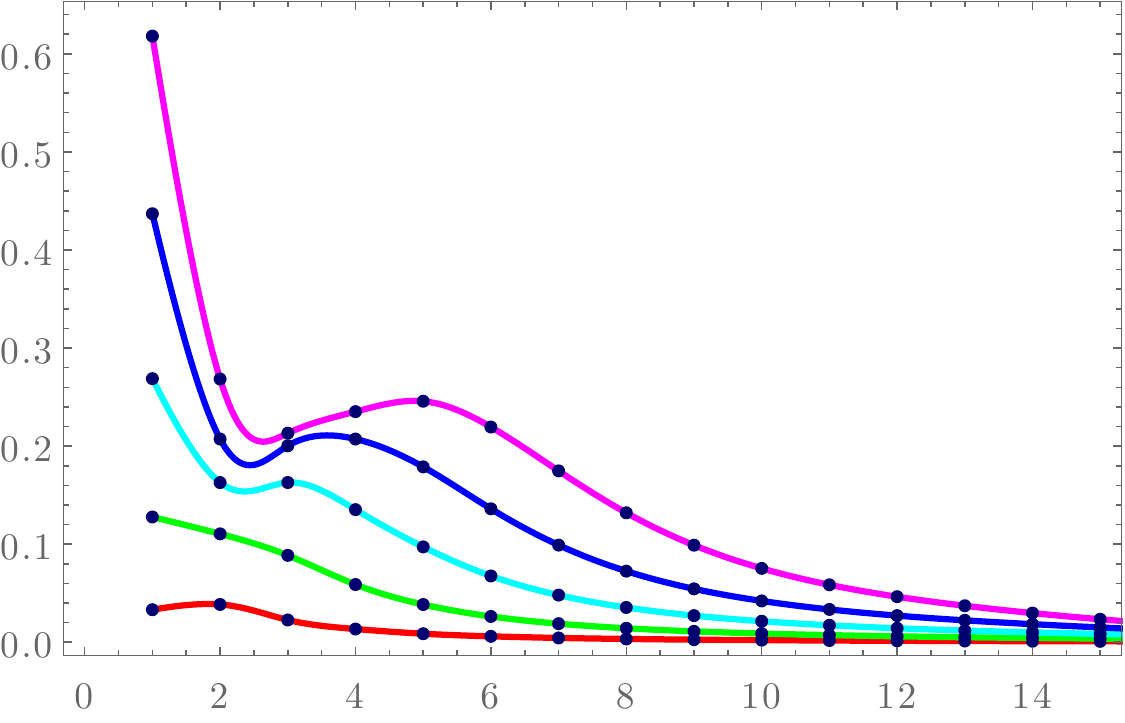}
		\caption{Number of particles in the $n$-th mode with no angular momentum created inside the disc by an instantaneous electric pulse that grows linearly from zero at the center to $eE$ at the boundary. The plots correspond to the values $eE=5,10,15,20,25$ from the bottom up ($m=R=1$). Interpolating curves (of order 3) are shown for clarity.}
		\label{disc}
			\end{minipage}
\end{figure}

As a straightforward application of this expression, we consider an instantaneous, radial electric field on a disc of radius $R$, whose magnitude grows linearly from zero at the center of the disc to a finite value $E$ at its boundary. The normal modes are characterized by the angular momentum $\ell$ and the number of radial nodes $n$. From \eqref{eq:number-irrot-discrete}, the number of particles created at the mode $(\ell,n)$ is easily seen to be
	\begin{align}
		N_{\ell,n}=\frac{1}{R^4}\ \sum_{n'}
		\frac{(\omega_{\ell n}-\omega_{\ell n'})^2}{\omega_{\ell n}\,\omega_{\ell n'}}
		\left|\int_0^R dr\,r
		\ \frac{J_\ell(j^{(\ell)}_n\,r/R)}{J_{\ell+1}(j^{(\ell)}_n)}
		\ \frac{J_{\ell}(j^{(\ell)}_{n'}\,r/R)}{J_{\ell+1}(j^{(\ell)}_{n'})}
		\ e^{\frac i2eE\,r^2}\right|^2\,,
	\end{align}	
where
\begin{align}
	\omega_{\ell n}=\sqrt{\frac{(j^{(\ell)}_n)^2}{R^2}+\frac{\ell^2}{R^2}+m^2}\,.
\end{align}
Figure \ref{disc} shows the number of produced particles as a function of the quantum number $n$, for zero angular momentum.

\section{Discussion and Conclusions}\label{sec:conclu}

In this article we have computed the rate of pair production driven by an electric field which is normal to the boundary. Although boundaries had already been incorporated in the context of particle creation by electric fields in many different scenarios, there is, up to our knowledge, no other example which, together with \cite{Martin:1988de,Martin:1988gr}, provides an explicit solution in the presence of a normal component of the field. This is a consequence of the fact that this orientation of the field causes technical difficulties which limit the use of standard methods, such as heat-kernel or worldline instantons.

\begin{figure}
	\centering
	\begin{minipage}{.4\linewidth}
		\includegraphics[width=\linewidth]{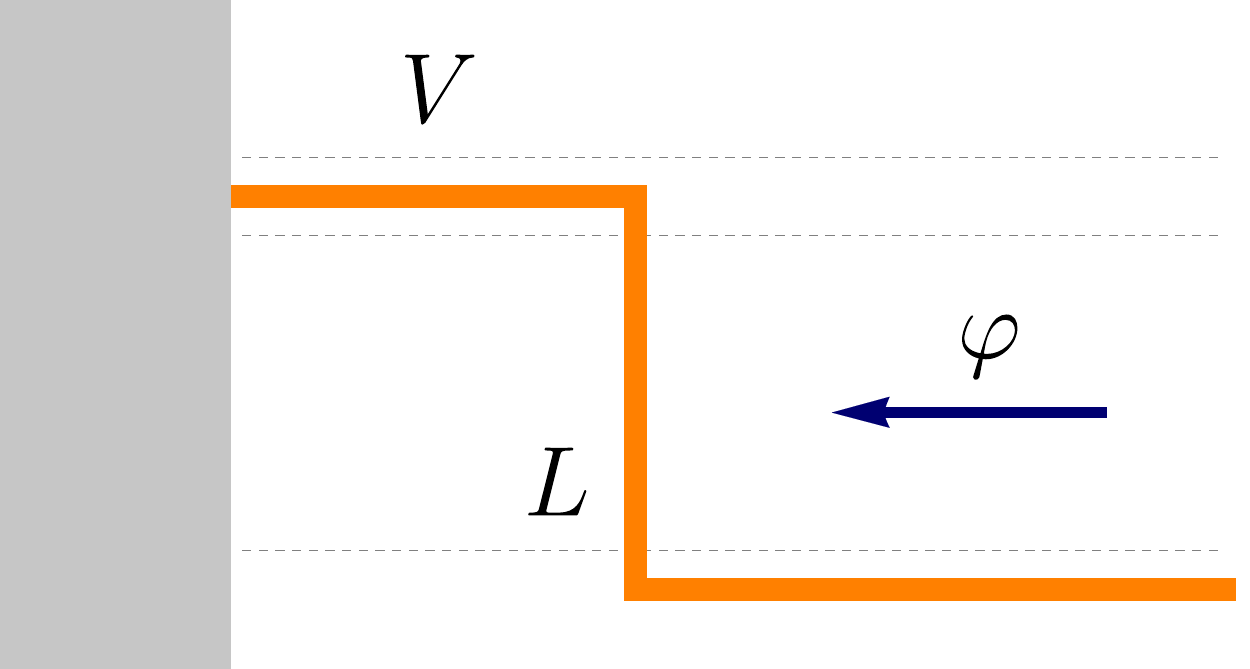}
		\caption{Particle in the half-line colliding against a step of height $V$ and width $L$. Dashed lines delimit the mass gaps.}
		\label{steplike}
	\end{minipage}\hspace{10mm}
	\begin{minipage}{.4\linewidth}
		\includegraphics[width=\linewidth]{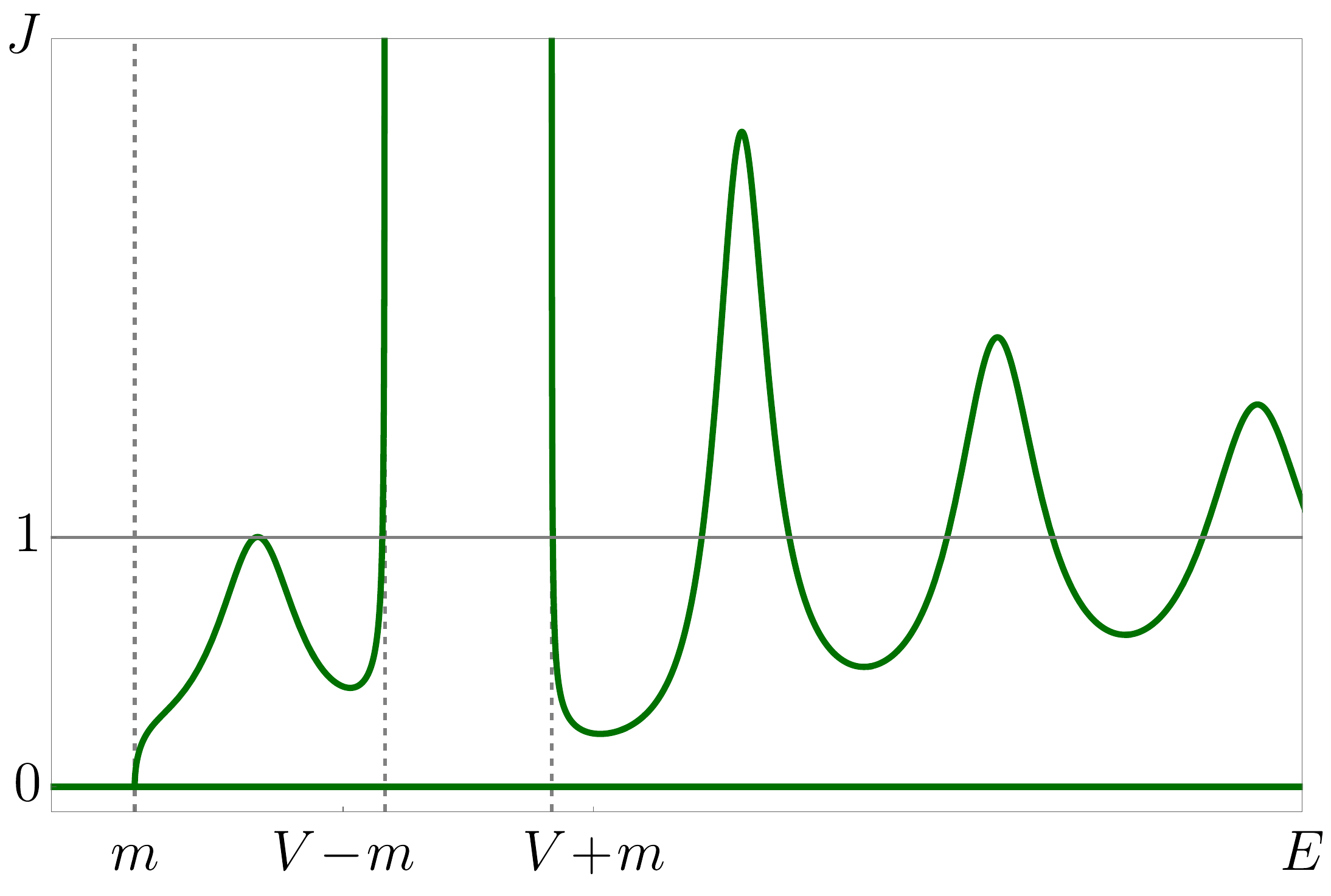}
		\caption{The bouncing current within the step. For energies in the Klein region $m<E<V-m$ this current represents propagating antiparticles.}
		\label{bounce}
	\end{minipage}
\end{figure}

At this point, it is worth mentioning an example of a static electric field normal to the boundary which admits an explicit solution: a one-dimensional step-like potential in the half-line. More precisely, let us consider a scalar field in the half-line subject to a scalar potential $A_0=V\,H(L-x)$, as shown in Figure \ref{steplike}.
With Dirichlet boundary conditions, the stationary solutions inside the step, $0<x<L$, take the form
\begin{align}\label{debajo}
	\varphi(x,t)=-\frac{2ip\,e^{-ipL}}{q\cos{(qL)}-ip\sin{(qL)}}\sin{(qx)}\,e^{-iEt}\,,
\end{align}
with $p=\sqrt{E^2-m^2}$ and $q=\sqrt{(E-V)^2-m^2}$. For energies in the so-called Klein zone, $m<E<V-m$, an incoming particle completely bounces at the step so the wave function at $x<0$ \eqref{debajo} must be interpreted as the result of infinitely many bounces within the step of antiparticles created by the delta-like electric field at $x=L$. From \ref{debajo}, one reads the left- and right-going antiparticle current $J$,
\begin{align}
	J=\frac{eqp}{q^2\cos^2{(qL)}+p^2\sin^2{(qL)}}\,,
\end{align}
which is represented in Figure \ref{bounce} for different values of the incoming energy. In the Klein zone, $m<E<V-m$, $J$ exclusively represents the current of the antiparticles. For higher energies, namely $E>V+m$, the incoming particle penetrates the barrier so $J$ contains this contribution as well.

Therefore, this static example in relativistic quantum mechanics contains pair creation by a normal electric field and displays the reflection of the created particles at the boundary. However, the phenomenon arises as an interpretation of the stationary states and not from a direct calculation of the rate of particle production.

In the setup considered in the present article, particles and antiparticles appear in a cleaner fashion. We have implemented as much simplifications as possible: we have considered a scalar field in the half-line subject to an instantaneous electric pulse modeled by a Dirac delta in time. The solution is obtained with the standard procedure of computing the Bogoliubov transformation.

The complexity entailed by the normal component of the field arises from the dynamics of the produced particles, which experience reflections at the boundary. To get some insight into this dynamics we solved first the electric pulse in the whole real line. We reproduce the already known result but we give the expression of the total density of produced particles, which, as far as we know, is not available in previous works. We also showed that the transferred energy diverges. More importantly for our purposes, this example lead us to a set of normal modes which is more appropriate for solving the problem in the half-line. In this way, the solution in the half-line can be interpreted as the result of two contributions: one arising from the reflections of particles produced at the bulk and another one which is a purely boundary contribution.

Subsequently, for comparison purposes, we also solved the case of an {instantaneous} electric pulse in a finite interval. The large-length limit of this case reproduces the half-line result. 

The effect of the boundaries is not only the restriction on the allowed momenta: the breaking of translation invariance manifests in the fact that the Bogoliubov transformations are no longer diagonal in momentum, giving rise to mode-mixing contributions to the particle density.

This analysis lead us to consider the case of an irrotational, {instantaneous} pulse in a finite region of space confined within a compact boundary. {We obtained an expression for the Bogoliubov coefficient $\beta$ where the dependence on the frequencies appears only in the frequency mismatch factor, which has a unique general expression in which the specific geometry is encoded in the matrix elements of the gauge phase. The expression is amenable for numerical evaluation, which we illustrate with the density of pair production in a disc driven by a radial electric pulse.}

\vspace{1cm}

\noindent{\bf Acknowledgments:} We thank support from CONICET (PIP 0262) and UNLP (I+D X1042 and X931).




\begin{thebibliography}{9}

\bibitem{Sauter}
F.~Sauter,
``Über das Verhalten eines Elektrons im homogenen elektrischen Feld nach der relativistischen Theorie Diracs,''
\href{https://doi.org/10.1007/BF01339461}{Z. Physik \textbf{69}, 742–764 (1931)}

\bibitem{H-E}
W.~Heisenberg, H.~Euler,
``Folgerungen aus der Diracschen Theorie des Positrons,''
\href{https://doi.org/10.1007/BF01343663}{Z. Physik \textbf{98}, 714-732 (1936)} (English translation in W. Korolevski, H. Kleinert, \href{https://arxiv.org/pdf/physics/0605038}{arXiv:physics/0605038})

\bibitem{Weisskopf:1936hya}
V.~Weisskopf,
``The electrodynamics of the vacuum based on the quantum theory of the electron,''
Kong. Dan. Vid. Sel. Mat. Fys. Med. \textbf{14N6} (1936) no.6, 1-39. In: A.~I.~Miller (ed.),
``Early quantum electrodynamics: A source book,''
\href{https://doi.org/10.1017/CBO9780511608223}{(1994) Cambridge University Press}

\bibitem{Schwinger:1951nm}
J.~S.~Schwinger,
``On gauge invariance and vacuum polarization,''
\href{https://doi.org/10.1103/PhysRev.82.664}{Phys. Rev. \textbf{82} (1951), 664-679}

\bibitem{Nikishov:1969}
A.~I.~Nikishov,
``Pair production by a constant external field,''
Zh. Eksp. Theor. Fiz. \textbf{57} (1969), 1210-1216
[\href{https://jetp.ras.ru/cgi-bin/dn/e_030_04_0660.pdf}{Sov. Phys.-JETP \textbf{30} (1970) 660}]

\bibitem{Gitman:1977ne}
D.~M.~Gitman,
``Processes of Arbitrary Order in Quantum Electrodynamics with a Pair Creating External Field,''
\href{https://doi.org/10.1088/0305-4470/10/11/026}{J. Phys. A \textbf{10} (1977), 2007-2020}.

\bibitem{Kluger:1991ib}
Y.~Kluger, J.~M.~Eisenberg, B.~Svetitsky, F.~Cooper and E.~Mottola,
``Pair production in a strong electric field,''
\href{https://doi.org/10.1103/PhysRevLett.67.2427}{Phys. Rev. Lett. \textbf{67} (1991), 2427-2430}.

\bibitem{Kluger:1992gb}
Y.~Kluger, J.~M.~Eisenberg, B.~Svetitsky, F.~Cooper and E.~Mottola,
``Fermion pair production in a strong electric field,''
\href{https://doi.org/10.1103/PhysRevD.45.4659}{Phys. Rev. D \textbf{45} (1992), 4659-4671}.


\bibitem{Smolyansky:1997fc}
S.~A.~Smolyansky, G.~Ropke, S.~M.~Schmidt, D.~Blaschke, V.~D.~Toneev and A.~V.~Prozorkevich,
``Dynamical derivation of a quantum kinetic equation for particle production in the Schwinger mechanism,''
\href{https://arxiv.org/pdf/hep-ph/9712377}{[arXiv:hep-ph/9712377].}

\bibitem{Schmidt:1998vi}
S.~M.~Schmidt, D.~Blaschke, G.~Ropke, S.~A.~Smolyansky, A.~V.~Prozorkevich and V.~D.~Toneev,
``A Quantum kinetic equation for particle production in the Schwinger mechanism,''
\href{https://doi.org/10.1142/S0218301398000403}{Int. J. Mod. Phys. E \textbf{7} (1998), 709-722},
\href{https://arxiv.org/pdf/hep-ph/9809227}{arXiv:hep-ph/9809227}.

\bibitem{Affleck:1981bma}
I.~K.~Affleck, O.~Alvarez and N.~S.~Manton,
``Pair Production at Strong Coupling in Weak External Fields,''
\href{https://doi.org/10.1016/0550-3213(82)90455-2}{Nucl. Phys. B \textbf{197} (1982), 509-519}

\bibitem{Dunne:2005sx}
G.~V.~Dunne and C.~Schubert,
``Worldline instantons and pair production in inhomogeneous fields,''
\href{https://doi.org/10.1103/PhysRevD.72.105004}{Phys. Rev. D \textbf{72} (2005), 105004},
\href{https://arxiv.org/pdf/hep-th/0507174}{arXiv:hep-th/0507174.}

\bibitem{Schubert:2024vft}
C.~Schubert,
``Pair creation in electric fields,''
\href{https://doi.org/10.1051/epjconf/202430103004}{EPJ Web Conf. \textbf{301} (2024), 03004},
\href{https://arxiv.org/pdf/2406.13135}{arXiv:2406.13135 [hep-ph]}.

\bibitem{Sauter2}
F.~Sauter, 
``Zum `Kleinschen Paradoxon''',
\href{https://doi.org/10.1007/BF01349862}{Z. Phys. 73, 547 (1932)}

\bibitem{Brezin:1970xf}
E.~Brezin and C.~Itzykson,
``Pair production in vacuum by an alternating field,''
\href{https://doi.org/10.1103/PhysRevD.2.1191}{Phys. Rev. D \textbf{2} (1970), 1191-1199}.

\bibitem{Popov:1971}
V.~S.~Popov,
``Pair production in a variable external field (quasiclassical approximation),''
Zh. Eksp. Theor. Fiz. \textbf{61} (1971), 1334-1351
[Sov. Phys.-JETP \textbf{34} (1971) 709].

\bibitem{Narozhnyi:1970uv}
N.~B.~Narozhnyi and A.~I.~Nikishov,
``The simplest processes in the pair creating electric field,''
Yad. Fiz. \textbf{11} (1970), 1072
[Sov.J.Nucl.Phys. \textbf{11} (1970) 596].

\bibitem{Hebenstreit:2010vz}
F.~Hebenstreit, R.~Alkofer and H.~Gies,
``Schwinger pair production in space and time-dependent electric fields: Relating the Wigner formalism to quantum kinetic theory,''
\href{https://doi.org/10.1103/PhysRevD.82.105026}{Phys. Rev. D \textbf{82} (2010), 105026},
\href{https://arxiv.org/pdf/1007.1099}{arXiv:1007.1099 [hep-ph]}

\bibitem{Beltran-Palau:2019bvz}
P.~Beltr{\'a}n-Palau, A.~Ferreiro, J.~Navarro-Salas and S.~Pla,
``Breaking of adiabatic invariance in the creation of particles by electromagnetic fields,''
\href{https://doi.org/10.1103/PhysRevD.100.085014}{Phys. Rev. D \textbf{100} (2019), 085014},
\href{https://arxiv.org/pdf/1905.07215}{arXiv:1905.07215 [hep-th]}.

\bibitem{Hebenstreit:2009km}
F.~Hebenstreit, R.~Alkofer, G.~V.~Dunne and H.~Gies,
``Momentum signatures for Schwinger pair production in short laser pulses with sub-cycle structure,''
\href{https://doi.org/10.1103/PhysRevLett.102.150404}{Phys. Rev. Lett. \textbf{102} (2009), 150404},
\href{https://arxiv.org/pdf/0901.2631}{arXiv:0901.2631 [hep-ph]}

\bibitem{Breev:2021lpn}
A.~I.~Breev, S.~P.~Gavrilov, D.~M.~Gitman and A.~A.~Shishmarev,
``Vacuum instability in time-dependent electric fields: New example of an exactly solvable case,''
\href{https://doi.org/10.1103/PhysRevD.104.076008}{Phys. Rev. D \textbf{104} (2021) no.7, 076008},
\href{https://arxiv.org/pdf/2106.06322}{arXiv:2106.06322 [hep-th]}

\bibitem{Schneider:2014mla}
C.~Schneider and R.~Sch{\"u}tzhold,
``Dynamically assisted Sauter-Schwinger effect in inhomogeneous electric fields,''
\href{https://doi.org/10.1007/JHEP02(2016)164}{JHEP \textbf{02} (2016), 164},
\href{https://arxiv.org/pdf/1407.3584}{arXiv:1407.3584 [hep-th].}

\bibitem{Aleksandrov:2016lxd}
I.~A.~Aleksandrov, G.~Plunien and V.~M.~Shabaev,
``Electron-positron pair production in external electric fields varying both in space and time,''
\href{https://doi.org/10.1103/PhysRevD.94.065024}{Phys. Rev. D \textbf{94} (2016) no.6, 065024},
\href{https://arxiv.org/pdf/1606.06313}{arXiv:1606.06313 [hep-th]}

\bibitem{Aleksandrov:2019ddt}
I.~A.~Aleksandrov and C.~Kohlf{\"u}rst,
``Pair production in temporally and spatially oscillating fields,''
\href{https://doi.org/10.1103/PhysRevD.101.096009}{Phys. Rev. D \textbf{101} (2020) no.9, 096009},
\href{https://arxiv.org/pdf/1912.09924}{arXiv:1912.09924 [hep-th]}

\bibitem{Dunne:2006st}
G.~V.~Dunne, Q.~h.~Wang, H.~Gies and C.~Schubert,
``Worldline instantons. II. The Fluctuation prefactor,''
\href{https://doi.org/10.1103/PhysRevD.73.065028}{Phys. Rev. D \textbf{73} (2006), 065028},
\href{https://arxiv.org/pdf/hep-th/0602176}{arXiv:hep-th/0602176 [hep-th]}.

\bibitem{Dunne:2022esi}
G.~V.~Dunne and Z.~Harris,
``Resurgence of the effective action in inhomogeneous fields,''
\href{https://doi.org/10.1103/PhysRevD.107.065003}{Phys. Rev. D \textbf{107} (2023) no.6, 065003},
\href{https://arxiv.org/pdf/2212.04599}{arXiv:2212.04599 [hep-th]}.

\bibitem{DegliEsposti:2022yqw}
G.~Degli Esposti and G.~Torgrimsson,
``Worldline instantons for the momentum spectrum of Schwinger pair production in spacetime dependent fields,''
\href{https://doi.org/10.1103/PhysRevD.107.056019}{Phys. Rev. D \textbf{107} (2023) no.5, 056019},
\href{https://arxiv.org/pdf/2212.11578}{arXiv:2212.11578 [hep-ph]}.

\bibitem{Wang:1988ct}
R.~C.~Wang and C.~Y.~Wong,
``Finite Size Effect in the Schwinger Particle Production Mechanism,''
\href{https://doi.org/10.1103/PhysRevD.38.348}{Phys. Rev. D \textbf{38} (1988), 348-359}

\bibitem{Wong:1994ei}
C.~Y.~Wong, R.~C.~Wang and J.~S.~Wu,
``Schwinger particle production mechanism for a finite length flux tube with transverse confinement,''
\href{https://doi.org/10.1103/PhysRevD.51.3940}{Phys. Rev. D \textbf{51} (1995), 3940-3943,}
\href{https://arxiv.org/pdf/hep-ph/9412357}{arXiv:hep-ph/9412357 [hep-ph]}.

\bibitem{Gavrilov:2015zem}
S.~P.~Gavrilov and D.~M.~Gitman,
``Scattering and pair creation by a constant electric field between two capacitor plates,''
\href{https://doi.org/10.1103/PhysRevD.93.045033}{Phys. Rev. D \textbf{93} (2016) no.4, 045033},
\href{https://arxiv.org/pdf/1511.02915}{arXiv:1511.02915 [hep-th]}

\bibitem{McGady:2017slr}
D.~A.~McGady and J.~Brog{\r{a}}rd,
``Short-distance Schwinger-mechanism and chiral symmetry,''
\href{https://arxiv.org/pdf/1711.03504}{arXiv:1711.03504 [hep-th]}

\bibitem{Breev:2019kug}
A.~I.~Breev, S.~P.~Gavrilov and D.~M.~Gitman,
``In- and out-states of scalar particles confined between two capacitor plates,''
\href{https://doi.org/10.33910/2687-153X-2020-1-1-30-39}{Complex Syst. \textbf{1} (2020), 30-39},
\href{https://arxiv.org/pdf/1909.06948}{arXiv:1909.06948 [hep-th]}

\bibitem{Taya:2020pkm}
H.~Taya,
``Mutual assistance between the Schwinger mechanism and the dynamical Casimir effect,''
\href{https://doi.org/10.1103/PhysRevResearch.2.023346}{Phys. Rev. Res. \textbf{2} (2020) no.2, 023346},
\href{https://arxiv.org/pdf/2003.12061}{arXiv:2003.12061 [hep-ph].}

\bibitem{Brown:2015kgj}
A.~R.~Brown,
``Schwinger pair production at nonzero temperatures or in compact directions,''
\href{https://doi.org/10.1103/PhysRevD.98.036008}{Phys. Rev. D \textbf{98} (2018) no.3, 036008},
\href{https://arxiv.org/pdf/1512.05716}{arXiv:1512.05716 [hep-th]}.

\bibitem{Yamada:2024aca}
Y.~Yamada,
``Kaluza{\textendash}Klein Schwinger Effect,''
\href{https://doi.org/10.1093/ptep/ptae124}{PTEP \textbf{2024} (2024) no.8, 083B09},
\href{https://arxiv.org/pdf/2403.13451}{arXiv:2403.13451 [hep-th]}.


\bibitem{Martin:1988de}
C.~Martin and D.~Vautherin,
``Finite Size Effects in Pair Production by an External Field,''
\href{https://doi.org/10.1103/PhysRevD.38.3593}{Phys. Rev. D \textbf{38} (1988), 3593-3595}.

\bibitem{Martin:1988gr}
C.~Martin and D.~Vautherin,
``Finite Size and Dynamical Effects in Pair Production by an External Field,''
\href{https://doi.org/10.1103/PhysRevD.40.1667}{Phys. Rev. D \textbf{40} (1989), 1667}.

\bibitem{Mahajan:2008ya}
G.~Mahajan,
``Particle creation in a time-dependent electric field revisited,''
\href{https://doi.org/10.1016/j.aop.2008.10.012}{Annals Phys. \textbf{324}, 361-370 (2009)},
\href{https://arxiv.org/abs/0805.4270}{arXiv:0805.4270 [hep-th]}

\bibitem{Ilderton:2021zej}
A.~Ilderton,
``Physics of adiabatic particle number in the Schwinger effect,''
\href{https://doi.org/10.1103/PhysRevD.105.016021}{Phys. Rev. D \textbf{105} (2022) no.1, 016021},
\href{https://arxiv.org/pdf/2108.13885}{arXiv:2108.13885 [hep-ph].}

\bibitem{Nikishov:1970br}
A.~I.~Nikishov,
``Barrier scattering in field theory removal of klein paradox,''
\href{https://doi.org/10.1016/0550-3213(70)90527-4}{Nucl. Phys. B \textbf{21} (1970), 346-358}

\bibitem{Cohen:2008wz}
T.~D.~Cohen and D.~A.~McGady,
``The Schwinger mechanism revisited,''
\href{https://doi.org/10.1103/PhysRevD.78.036008}{Phys. Rev. D \textbf{78} (2008), 036008},
\href{https://arxiv.org/pdf/0807.1117}{arXiv:0807.1117 [hep-ph]}

\bibitem{Schenfeld:1990yu}
T.~Schenfeld, B.~Muller, K.~Sailer, J.~Reinhardt, W.~Greiner and A.~Schafer,
``Pair creation in a flux tube,''
\href{https://doi.org/10.1016/0370-2693(90)91039-E}{Phys. Lett. B \textbf{247} (1990), 5-12}.

\end{thebibliography}
\end{document}